\documentclass{webofc}
\usepackage[varg]{txfonts}   
\begin{document}
\title{The mm-to-cm SED of spiral galaxies}
\subtitle{Synergies between NIKA2 and SRT instruments}
%
%

\author{\firstname{Simone} \lastname{Bianchi}\inst{1}\fnsep\thanks{\email{simone.bianchi@inaf.it}} \and 
	\firstname{Matteo} \lastname{Murgia}\inst{2} 
	\and
	\firstname{Andrea} \lastname{Melis}\inst{2}
        \and
        \firstname{Viviana} \lastname{Casasola}\inst{3}
                                        \and
        \firstname{Maud} \lastname{Galametz}\inst{4}
                                \and
        \firstname{Fr\'ed\'eric} \lastname{Galliano}\inst{4}
        \and
        \firstname{Federica} \lastname{Govoni}\inst{2}
                        \and
        \firstname{Anthony} \lastname{Jones}\inst{5}
                \and
        \firstname{Suzanne} \lastname{Madden}\inst{4}
                        \and
        \firstname{Rosita} \lastname{Paladino}\inst{3}
                                \and
        \firstname{Emmanuel} \lastname{Xilouris}\inst{6}
                        \and
        \firstname{Nathalie} \lastname{Ysard}\inst{5}
}

\institute{INAF-Osservatorio Astrofisico di Arcetri, L. E. Fermi 50, 50125, Firenze, Italy
\and
            INAF-Osservatorio Astronomico di Cagliari, Via della Scienza 5, 09047 Selargius (CA), Italy
\and
            INAF-Istituto di Radioastronomia-Via P. Gobetti, 101, 40129 Bologna, Italy
           \and 
           AIM, CEA, CNRS, Universit\'e Paris-Saclay, Universit\'e  Paris Diderot, Sorbonne Paris Cit\'e, 91191 Gif-sur-Yvette, France
           \and
           Universit\'e  Paris-Saclay, CNRS, Institut d'Astrophysique Spatiale, 91405 Orsay, France
           \and
          National Observatory of Athens, Institute for Astronomy, Astrophysics, Space Applications and Remote Sensing, Ioannou Metaxa
and Vasileos Pavlou, 15236 Athens, Greece
          }

\abstract{%
The mm-to-cm range of the Spectral Energy Distribution of spiral galaxies remains largely unexplored. Its coverage is required to disentangle the contribution of dust emission, free-free and synchrotron radiation and can provide constraints on dust models, star-formation rates and ISM properties. We present the case for a synergy between NIKA2 observations of nearby spirals and those from planned and current instrumentation at the Sardinia Radio Telescope, and report on a pilot K-band program to search for Anomalous Microwave Emission, an elusive emission component which is presumably related to dust.
}
\maketitle
\section{Galactic SEDs in the mm-to-cm range}
\label{sec:intro}

\begin{figure}[h]
\begin{center}
\includegraphics[width=\hsize,trim= 13 6 1 15,clip]{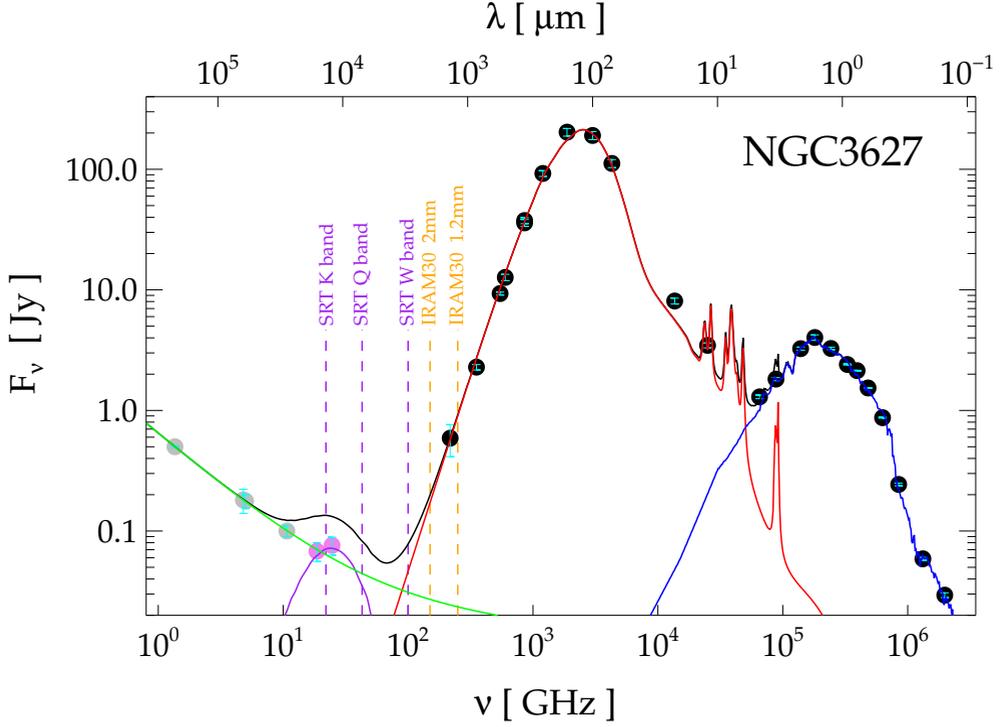}
\caption{The SED of NGC3627. The total SED (black) results from the sum of: 1) CIGALE fits of the stellar (blue line) and dust (red line) components to the DustPedia datapoints (black dots; \cite{NersesianA&A2019,ClarkA&A2018}); 2) synchrotron and free-free radiation fits to the mid-radio continuum emission (green line and grey dots; \cite{TabatabaeiApJ2017});
3) a prediction for AME (purple line), obtained using a model spectral template \cite{BattistelliApJL2019} normalised to the MW AME(30 GHz)/IR(100 $\mu$m) ratio \cite{PlanckA&A2016}. 
NIKA2 and the SRT instrumentation can cover the mm-to-cm gap. DustPedia datapoints at mm wavelengths are from (unresolved) {\em Planck} observations,
while the purple datapoints show the results from our pilot SRT program in the K band.
DustPedia images and photometry are available at {\tt http://dustpedia.astro.noa.gr/}.}
\label{fig:sed}       
\end{center}
\end{figure}

Our knowledge of the Spectral Energy Distribution (SED) of nearby galaxies has greatly improved in  recent times, thanks to multi-wavelength surveys and in particular to the  
{\em Herschel} Space Observatory. {\em Herschel} has extended the SED coverage to far-infrared (FIR) and submm wavelengths, a range where {\em dust}, i.e.\ solid grains 
in the interstellar medium (ISM), emits the radiation it has absorbed from starlight ($\sim$1/4 of the total luminosity for late type galaxies; \cite{BianchiA&A2018}). Using a dust  
model and fitting tools such as, e.g., CIGALE \cite{BoquienA&A2019} and HerBIE \cite{GallianoMNRAS2018}, a wealth of information can be extracted from the SED: star-formation rates (SFRs), stellar and dust masses, gas-to-dust ratios, etc. (\cite{NersesianA&A2019,AnianoApJ2020,DeLoozeMNRAS2020,GallianoA&A2021}, just to mention a few  recents works).
Figure~\ref{fig:sed} shows an example of the current SED knowledge: it uses data from {\em DustPedia} \cite{DaviesPASP2017}, a database collecting photometry and imagery for all the large ($D_{25} > 1'$) and nearby  ($d < 30$ Mpc) galaxies observed by {\em Herschel}; and mid radio continuum flux densities from \cite{TabatabaeiApJ2017}.

A few gaps still remain: in particular, data for $15\le \nu/\mbox{GHz} \le300$ (mm-to-cm range) is limited to the Milky Way (MW) and a few bright objects. 
The high-frequency side of this gap, at $\nu \approx 200$ GHz, is still dominated by dust emission: radiation could come from the Rayleigh-Jeans tail of
the warmer dust, or be due to colder grains undetected from FIR-submm observations. Large masses of cold grains, and/or variations
of the dust absorption cross-section at longer wavelengths could be responsible for the {\em submm excess} detected in a few objects \cite{GallianoRev2018}.
Thus, observations at these frequencies are important to constrain the properties of dust, which are still uncertain. Current dust models
are made to reproduce the observations of extinction, emission and depletion at high galactic latitude in the MW \cite{CompiegneA&A2011}.
Various dust models exist, leading to different estimates of quantities such as the dust mass (see, e.g., \cite{NersesianA&A2019}).
Furthermore, recent lab measurements are questioning the validity of the optical properties generally assumed for the grain materials \cite{DemykA&A2017}.
Finally, there are indications that the bulk of dust in a galaxy might have different properties than those of the diffuse MW medium \cite{BianchiA&A2019,ClarkMNRAS2019}.
The NIKA2 camera at the IRAM 30m radiotelescope,  and in particular the IMEGIN guaranteed-time project observing large nearby galaxies
at millimetre wavelengths \cite{Katsioli,Ejlali}, will help in shedding light on these issues.

At lower frequencies, emission components other than dust need to be taken into account. Already in NIKA2 maps at 2~mm, about 10\% of the SED is 
due to free-free and synchrotron radiation; their contribution increases, and that of dust reduces, in the mid and low-frequency side of the $15\le \nu/\mbox{GHz} \le300$ gap
(see, e.g.. \cite{Katsioli}). Free-free emission gives the maximum contribution in the mid of the frequency gap. Being due mostly to free electrons, this radiation can be directly related to 
ionising photons and thus provides a direct estimate of the star-formation rate, unaffected by the dust extinction corrections needed for estimates based on the UV continuum and 
the hydrogen  recombination lines in the optical \cite{MurphyProc2018}; or by the uncertain fraction of dust heating due to young stars, for estimates based on infrared
dust emission \cite{NersesianA&A2019}. The dominant emission toward the low-frequency side of the gap is due to synchrotron radiation, produced by cosmic ray electrons accelerated 
to relativistic velocities in supernova remnants and spiralling along the magnetic field in the ISM \cite{MurphyProc2018}. Since all of these emission mechanisms overlap
in frequency, full SED coverage is needed to solve the degeneracies and estimate the exact contribution of each component. 

An opportunity to extend the knowledge of the SED beyond the frequencies accessible to the NIKA2 camera is soon going to be offered at the Sardinia Radio telescope (SRT \cite{PrandoniA&A2017}; see Fig.~\ref{fig:sed}). In the incoming year, the MISTRAL camera, operating in the W Band at $\sim   90$~GHz, will be installed at SRT 
\cite{DAlessandro}, as well as a 19-feed Q band receiver at $\sim 40$~GHz \cite{OrfeiProc2020}. Thanks to the large SRT antenna, 64m, these instruments will allow us to map 
nearby galaxies with HPBW=12$''$ and 28$''$, respectively, resolutions close to those of NIKA2 and the {\em Herschel} instruments. Already available,  a 7-feed spectro-polarimetric 
receiver allows access to the K~band at $\sim 22$~GHz \cite{OrfeiProc2010}.

\section{Anomalous Microwave Emission}
\label{sec:ame}

A further emission component might need to be considered: the Anomalous Microwave Emission (AME). First detected in the MW in the mid-90s by Cosmic Microwave Background experiments, AME is seen as an excess over the expected free-free and synchrotron emissions, peaking at $\approx 30$ GHz and correlating with dust emission at larger frequencies (see \cite{DickinsonRev2018} for a review). AME is seen both in isolated clouds \cite{PlanckA&A2014},
as well as in the diffuse MW medium  where its amplitude, relative to the 100~$\mu$m surface brightness, is AME(30~GHz)/I(100~$\mu$m) $\approx$ 1/3000 \cite{PlanckA&A2016}.
 
The most accredited theory explains AME with electric dipole emission from rapidly rotating grains \cite{DraineApJ1998a}. For typical  ISM conditions, these grains must have size of $10^{-9}$ m, in order to spin fast and emit at the AME frequencies: they could be associated with the macromolecules carriers of the PAH features in the mid-infrared spectrum \cite{DraineApJ1998b}, or with nanosilicates \cite{HensleyApJ2017}. Alternatively, AME could be due to an
enhancement of the absorption/emission cross sections at microwave frequencies, either due to magnetic inclusions in grains \cite{DraineApJ1999} or  the amorphous state of the grain material 
\cite{JonesA&A2009,NashimotoPASJ2020}.
AME can thus set constraints  on the grain size distribution (and ISM conditions), under the spinning grain hypothesis, and/or on the grain composition, for the thermal emission hypothesis. 
Though these models can all explain AME, yet there is no concluding evidence for  its physical mechanism:  some authors find a good correlation with small-grain emission, thus 
favouring spinning dust  \cite{YsardA&A2010,HarperMNRAS2015,BellPASJ2019}; others support the thermal hypothesis, reporting a better correlation with the total dust emission \cite{HensleyApJ2016}.

While AME is almost ubiquitous in the MW, it has been detected with high significance in three extragalactic sources only. AME has been found with pointed observations in just one star-forming region in NGC~6946 \cite{MurphyApJ2010} and one compact radio-source in NGC~4725 (\cite{MurphyApJ2018}, even though, in this case, it does not appear to be associated with dust \cite{MurphyApJ2020}). On the global SED, strong evidence for AME has been recently found in the Andromeda galaxy \cite{BattistelliApJL2019}, with an AME component compatible with that in the MW.

\section{A pilot K-band program at SRT}
\label{sec:pilot}

As a preliminary project, at the beginning of 2021 we observed  four galaxies at SRT in the K band. They are: NGC~3627 (M~66), NGC~4254 (M~99), NGC4736 (M~94) and NGC5055 (M63).
These galaxies are part of the NIKA2 IMEGIN sample, of the DustPedia database and of the KINGFISHER project. DustPedia  provides homogenised images, full UV-to-submm global photometry and SED models \cite{ClarkA&A2018,NersesianA&A2019}, while KINGFISHER provides radio continuum data and models at frequencies between 1 and 10 GHz \cite{TabatabaeiApJ2017}. The first objective of the observations was the detection of AME, by comparing the SRT data with the available fits for the free-free and synchrotron components. Ultimately, the targets could become the first nucleus for a study of
the complete UV-to-radio SED, once the NIKA2 data and the future SRT instruments  cover the frequency gap between the submm and the K band (Fig.~\ref{fig:sed}).

The K-band receiver was used together with the SArdinia Roach2-based Digital Architecture for Radio Astronomy back end (SARDARA; \cite{MelisJAInst2018}). We selected 
two 1.2 GHz sub-bands at the edges of the K-band, at 18.6 and 24.6~GHz (with HPBW = 57$''$ and 45$''$, respectively). These sub-bands were covered by 1.46~MHz frequency 
channels in full Stokes mode. Observations were done on-the-fly over an area of $16'\times16'$, centered on each galaxy, with a scan speed 2$'$/s and scan separation HPBW/3. 
The area was alternatively covered by scanning along orthogonal RA and Dec directions, for an efficient removal of the scan noise. Each coverage took about 20 minutes 
and was repeated 15 to 40 times, depending on the target, frequency and observing conditions.

\begin{figure}[ht]
\begin{center}
\includegraphics[width=\hsize]{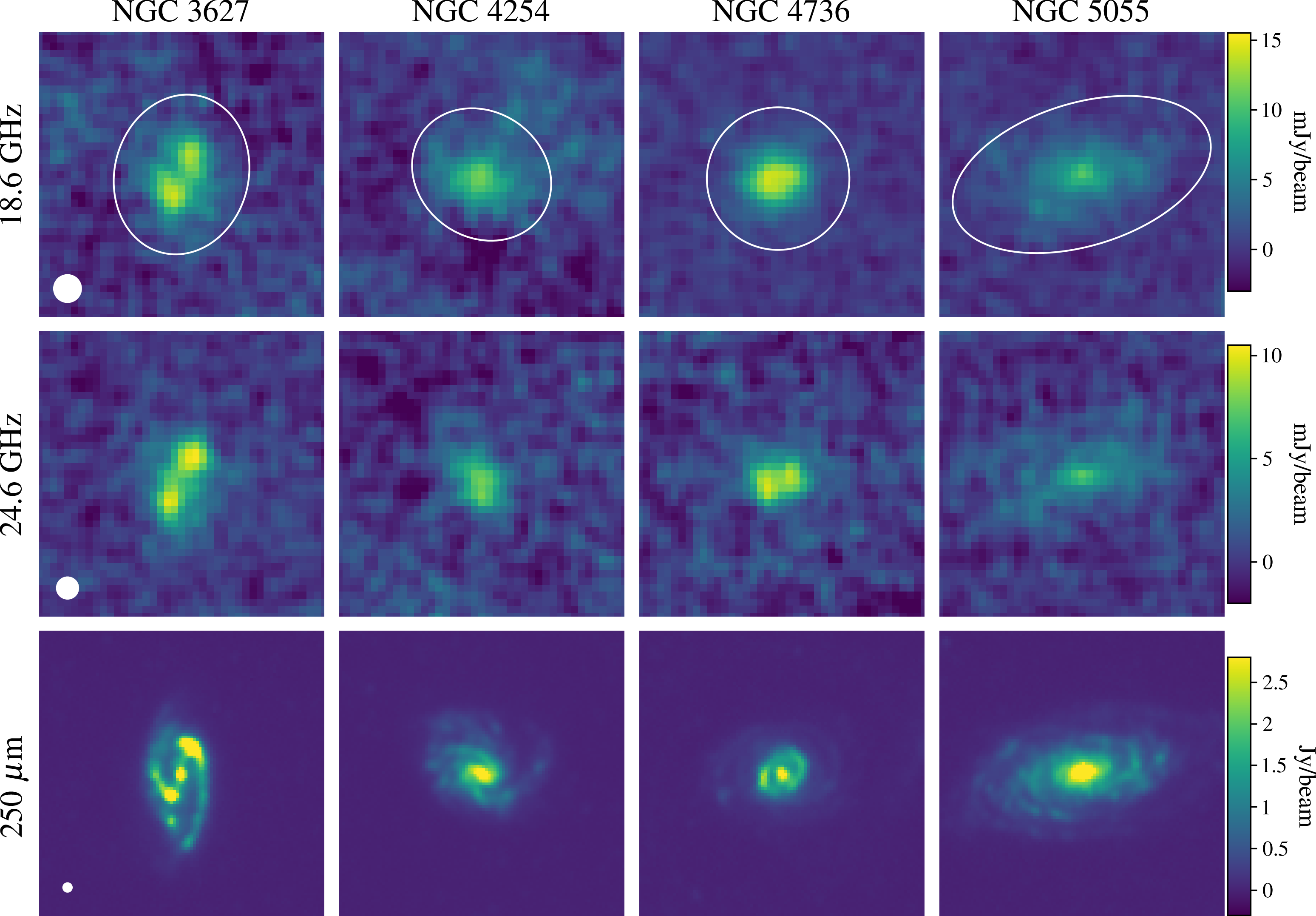}
\caption{Total intensity maps for our targets  (top and middle panels). For reference,  250~$\mu$m {\em Herschel}-SPIRE maps from DustPedia are shown in the bottom panels. We also show the areas used for photometry and the HPBW. Each panel is $10'\times10'$.}
\label{fig:images}       
\end{center}
\end{figure}

The spectral cubes were reduced using the proprietary Single-dish Spectral-polarimetry Software (SCUBE; \cite{MurgiaMNRAS2016})  that accounts for calibration, RFIs removal, baseline removal and map-making.  Maps for total and polarised intensity were made with pixel size 15$''$. Total intensity maps are shown in Fig.~\ref{fig:images}. Even though some 
mechanisms proposed for AME are expected to produce polarised emission \cite{DickinsonRev2018}, we detected none.

\section{Preliminary results}
\label{sec:results}

Using the apertures shown in Fig.~\ref{fig:images}, we measured the global flux density for our targets at 18.6~GHz and  24.6~GHz. These new datapoints were analized together 
with the mid-radio continuum from KINGFISHER. For NGC~4254 and NGC~4736, the SRT observations are consistent with the synchrotron and free-free fits to data for $\nu \le 10$~GHz. For NGC~3627  and NGC~5055, the 18.6~GHz flux density is still compatible with the fit in \cite{TabatabaeiApJ2017}; the 24.6~GHz
flux density, instead, is in excess of the fit extrapolation and larger than that at 18.6~GHz (see Fig.~\ref{fig:sed} for NGC~3627).

With the current data, it is difficult to asses if the 24.6~GHz excess seen in NGC~3627  and NGC~5055 is due to an additional component, such as AME, or if it is just a statistical fluctuation. Observations at larger frequencies, like those that will be possible with the forthcoming Q-band receiver for SRT, are needed to confirm this finding.
In fact, a model including synchrotron and free-free only can be fit to the entire $\nu\le 24.6$~Ghz range, resulting in minor modifications with respect to the fit for $\nu \le 10$~GHz.
The new parameters for the two radio components are still consistent with the spread  of properties found in the KINGFISHER sample. From the free-free component, estimates of the
global SFR can be obtained, which are consistent with those from the UV-to-submm SED fitting, though affected by large uncertainties \cite{Bianchi2022}. Again, more precise determinations
will need to wait for data at higher frequencies, or better to the full coverage of the $15\le \nu/\mbox{GHz} \le300$ gap, for a simultaneous analysis of all the emission components, dust included.

Currently, we can only put upper limits on the AME contribution at 30~GHz. We find that AME(30~GHz)/I(100~$\mu$m) is at least a factor 5 smaller that the values measured in M~31 and the MW (see the comparison between predicted  AME and SRT observations in Fig.~\ref{fig:sed}). This does not necessarily imply that the AME properties are different. In fact, the 
AME(30~GHz)/I(100~$\mu$m) ratio is an ill-defined characterisation of the AME emissivity, as it is expected to be lower in objects where the dust temperature is higher
\cite{TibbsProc2012}. Indeed, the average dust temperatures of our galaxies, as well as those in three other targets with no AME detection \cite{PeelMNRAS2011} are higher
than those of M~31 and the MW. When the AME emissivity is derived in terms of the column density of dust, available from the DustPedia analysis, the upper limits in our galaxies
become consistent with the AME detections. Apparently, AME may not be detected in our targets and those in \cite{PeelMNRAS2011} because of their relatively 
stronger radio-continuum than in the Milky Way and Andromeda galaxy \cite{Bianchi2022}. Future searches for AME might need to consider this in the selection of targets.

\section*{Acknowledgements}
We acknowledge support from the INAF mainstream 2018 program "Gas-DustPedia: A definitive view of the ISM in the Local Universe" and from grant PRIN MIUR 2017- 20173ML3WW\_001.

%
%
%

\end{document}